\documentclass[reprint,prb,superscriptaddress,showpacs]{revtex4-2}%
\usepackage{graphicx}
\usepackage{siunitx}
\usepackage{color}
\usepackage{amsmath}
\usepackage{amsfonts}
\usepackage{amssymb}
\usepackage{framed}
\usepackage{float}

\providecommand{\U}[1]{\protect\rule{.1in}{.1in}}
\makeatletter
\renewcommand*{\fnum@figure}{{\normalfont\bfseries \figurename\thefigure}}
\renewcommand*{\@caption@fignum@sep}{\textbf{ : }}
\makeatother
\usepackage{hyperref}
\hypersetup{
    colorlinks=true,
    linkcolor=blue,
    filecolor=magenta,      
    urlcolor=cyan,
}
\makeatletter

\DeclareTextSymbolDefault{\textquotedbl}{T1}

\usepackage{dcolumn}
\usepackage{bm}

\newcommand{\BNB}{Ba$_{14}$MnBi$_{11}$}

\makeatother

\begin{document}
\title{The mysterious magnetic ground state of Ba$_{14}$MnBi$_{11}$ is likely
altermagnetic}
\author{Po-Hao Chang}
\email{pchang8@gmu.edu}

\author{Igor I. Mazin}
\email{imazin2@gmu.edu}

\affiliation{Department of Physics and Astronomy, George Mason University, Fairfax,
VA 22030, USA}
\affiliation{Quantum Science and Engineering Center, George Mason University, Fairfax,
VA 22030, USA}
\begin{abstract}

Mn-based transition metal Zintl compounds in the ``14-1-11''
phase are known to host complex atomic and magnetic structures owing
to their intricate crystal structure. Among this family
of compounds, \BNB\ stands out as one of the least
understood compounds, with experimental measurements
and theoretical findings largely inconsistent. Following
up on the earlier attempt [D. S\'{a}nchez-Portal et al., Phys. Rev. B 65, 144414 (2002)] at establishing
a connection between metallicity and magnetism through a DFT-based
analysis, our work aims to provide additional insights to resolve the
existing contradictions. Our key findings is that the magnetic
ground state is very susceptible to charge doping (band filling).
Density functional calculations for stoichiometric Ba$_{14}$MnBi$_{11}$ give a rather stable ferromagnetic metallic ground state.
However, by adding
exactly one additional electron per Mn, the system becomes semiconducting
as expected, and consequently the magnetic ground state becomes weakly
antiferromagnetic (AF). On the other hand, upon small
hole doping the system transitions to a special type of AF state
known as altermagnetic ordering. The observed trends suggest that
hole and electron doping-induced phase transitions likely result from
different underlying mechanisms, influencing various exchange pathways.
Additionally, our projected density-of-states along with bandstructure
analyses indicate that, besides the largest hole contribution coming
from the tetrahedral unit of Bi, the isolated Bi sites also play a
substantial role and the dispersive bands near VBM suggest a rather
complex hybridization network involving both Bi band characters.
Through a comprehensive comparison of available data and our analysis,
we propose that the inconsistency in magnetic states between experimental
findings and DFT calculations is not a failure of DFT, but rather indicates nonstoichiometric effects,  likely impurities or defects, contributing
to the observed discrepancies. A possibility of stabilizing, through doping, an altermagnetic state, is exciting.
\end{abstract}
\maketitle

\section{Introduction}

Zintl compounds with the general formula A$_{14}$MnPn$_{11}$, where A is an alkaline earth, an alkaline metal, or a trivalent rare earth, and Pn is a pnictogen, have been attracting attention as potential thermoelectrics (especially Yb$_{14}$MnPn$_{11}$) for a few decades. These were intensively studied in 1992-2002, but larger forgotten after that. Yet, their physics, and especially their magnetic properties remain very poorly understood, which is especially true about \BNB, as summarized in Ref. \cite{sanchez-portal_bonding_2002}. 
Below we list the most enigmatic experimental and computational findings \cite{webb_new_1991,kuromoto_structure_1992,sanchez-portal_bonding_2002,miller_magnetism_2004} 

1. Magnetic susceptibility is clearly antiferromagnetic (AF), but DFT calculations, with large confidence, give ferromagnetism (FM) as the ground state.

2. Yet, the Curie-Weiss temperature is positive (ferromagnetic). This is a common phenomenon in A-type antiferromagnets consisting of strongly coupled FM planes, with a weak AF interaction between them. However, the crystal structure of \BNB, in its space group I4$_1$acd, is rather three-dimensional and, at first glance, highly symmetric: each Mn has 10 nearest neighbors in all directions, with four Mn-Mn bonds forming a tetrahedron, another four exactly the same tetrahedron, but flipped, and two more forming a linear structure. Geometrically, it seems that the structure cannot be easily partitioned into weakly coupled FM sublattices. 

3. The Curie-Weiss effective magnetic moment is 4.8 $\mu_B$, consistent with the spin $S\approx 2$ ($M=4$ $\mu_B$), but the saturation magnetization was claimed from the experiment in the field up to 5 T is $\approx 3$ $\mu_B$. At the same time, DFT calculations, again with a great degree of confidence show Mn to be in a divalent high-spin state, i.e., $M=5$ $\mu_B$.

4. Calculations give a metallic ground state, with a few Bi-derived bands at the Fermi level, and no Mn bands near $E_F$, consistent with Mn$^{2+}$. It was claimed to be metallic in the experiment as well, but in the only transport measurement \cite{kuromoto_structure_1992} the resistivity is barely dependent on temperature, with Residual Resistivity Ratio (RRR) about 1.3, and the resistivity being weakly metallic below the Ne\'el transition. 


5. Last but not least, the linear specific heat coefficient corresponds to the density of states of 44 states/eV$\cdot$formula, while the calculations give, for a sizeable range of Fermi energies, less than 7.5 states/eV$\cdot$formula.

So, in and by itself it is an intriguing compound with many unresolved mysteries. In addition, there is no experimental data on the magnetic pattern, except that it is clearly antiferromagnetic. As pointed out in Ref. \cite{sanchez-portal_bonding_2002}, there are three distinct-symmetry magnetic patterns even for $q=0$, which they called AFM(a-c). In retrospect, one of these patterns, namely AFM(b) is {\it altermagnetic} (AM), a novel type of ordered collinear magnetism that has been intensively discussed in the last few years\cite{mazin_altermagnetism_2023,mazin_editorial_2022,smejkal_emerging_2022,smejkal_beyond_2022}

In this paper, we will address some of the unresolved issues
either exact or speculative, assuming that the experimentally available samples have a small hole doping of the order of 1 hole per 4 formula units (F.U.), for instance, due to 1\% vacancies on Ba site. Interestingly, at the same time, doping stabilized the AM order.
{The paper is organized as follows. }Section \ref{sec:structure}
provides an overview of the crystal structure. The computational
details are described in Section \ref{sec:method}.
Section \ref{sec:edope} presents our data for the doping effect on
the ground states and how two different AF orderings can be induced.
Section \ref{sec:jdope} discusses the doping effect on the exchange
coupling using two different methods. Section \ref{sec:jpath} analyzes
the band structure and the projected density-of-states near the Fermi energy to establish a connection between band characters and
the doping effect on the magnetic structure. Finally, the conclusions
are summarized in Section \ref{sec:conclusion} 

\section{Crystal structure\protect\label{sec:structure}}

\begin{figure*}
\centerline{\includegraphics[scale=0.40]{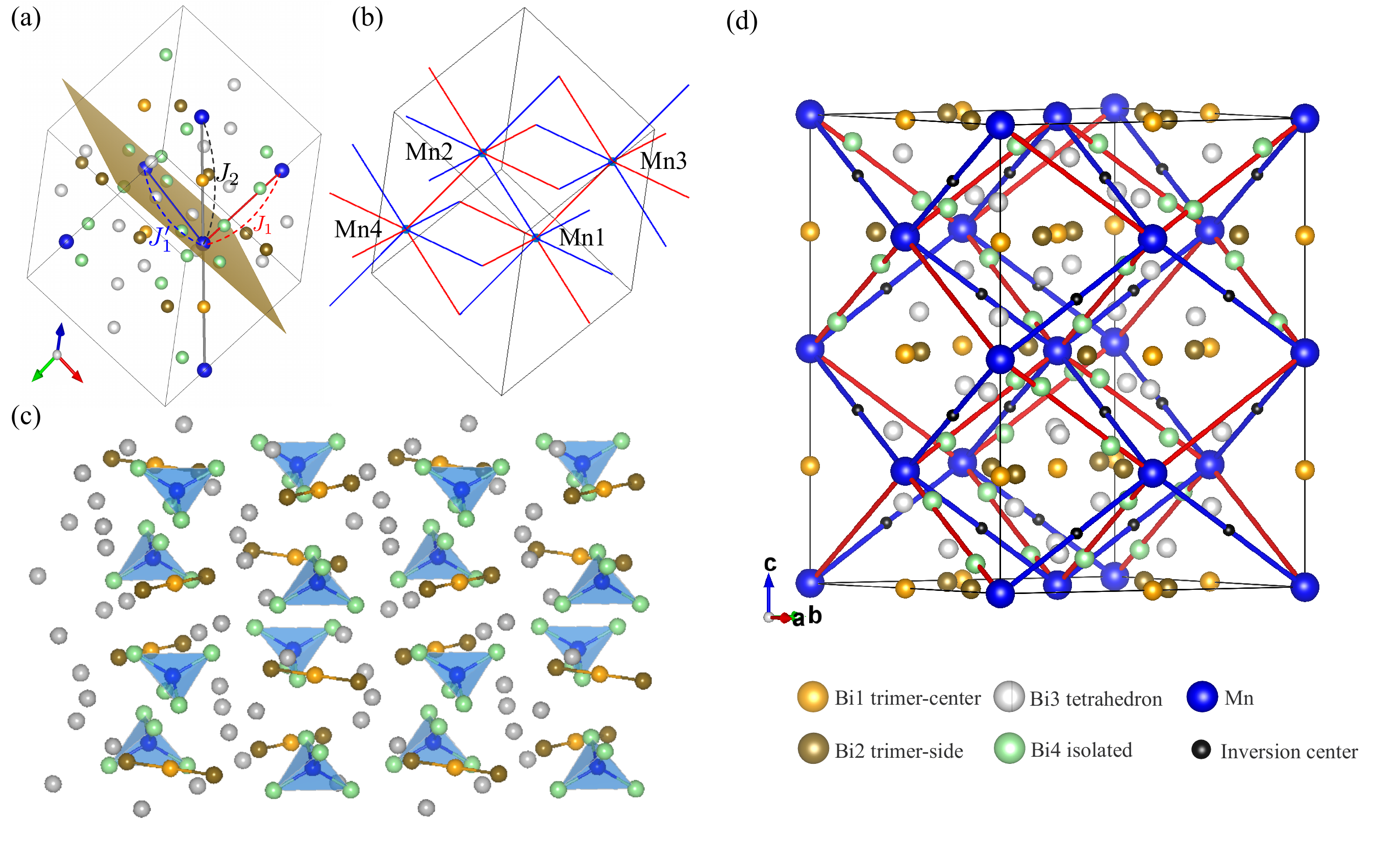}}
\caption{\protect\label{fig:structure} 
The crystal structure views with different emphases:
(a) The schematic view of the first three nearest exchange paths defined in the unitcell of four formula unit.
(b) Depiction of the model magnetic cell showing four magnetic sites and how two inequivalent exchange coupling constants 
$J_1$ and $J_{1'}$ are spatially oriented.
(c) 
Extended structural view revealing four inequivalent Bi sites organized into tetrahedral units, trimers, and isolated Bi sites.
(d) Additional symmetry feature: The view illustrates that $J_{1}$ bonds have an inversion center at the midpoints and $J_{1'}$ do not. 
}
\end{figure*}

The compound has a body-center-tetragonal space group I4$_1$acd unit cell with 
four formula units as depicted
in Fig. \ref{fig:structure} (a) with additional Mn sites (blue) included to show $J_2$.
There are four inequivalent Bi sites shown in different colors as defined in the figure.
The lattice parameters $a=18.665(3)$ and $c=24.429$ {\AA} are taken from the original 
experimental work \cite{kuromoto_structure_1992}.

While all Mn are symmetry equivalent, there are two inequivalent 1st NN exchange coupling
interactions: $J_{1}$ (red) that passes through two vertices from two tetrahedra
and $J_{1'}$ (blue) that do not. More importantly, as Fig. \ref{fig:structure} (d) shows, the  $J_{1}$ bonds have an inversion center at the midpoints and $J_{1'}$ do not. 

A more complete view of how $J_{1}$  and $J_{1'}$ are spatially oriented for each Mn site is presented in Fig. \ref{fig:structure} (b).  All four Mn sites are labeled with numbers to indicate their order for later discussion.

Fig. \ref{fig:structure} (c) shows the extended view of the structure
that exhibits a complicated tetrahedra network. Due to the large number
of atoms in the unit cell, for better visibility, Ba sites are not
displayed. It becomes clear that the crystal structure per F.U.
consists of one linear trimer (Bi$_{3}^{7-}$), four isolated Bi$^{3-}$
and a tetrahedral unit (MnBi$_{4}^{9-}$). The tetrahedra, as explained
in Ref. \cite{sanchez-portal_bonding_2002}, are interconnected through either $J_{1}$
and $J_{1}'$, two inequivalent first NN exchange coupling interactions,
which form two different networks. 

\section{Methods\protect\label{sec:method}}

The small energy scale of the magnetic interactions and the
large crystal structure require considerable computational flexibility; since different DFT codes are best at addressing different aspects of the calculations, we employed three DFT codes
of different types of basis sets to ensure accuracy as well as provide more
insight. As a reference, the electronic structure calculations were
first performed to obtain the total energies using Vienna ab initio
Simulation Package (VASP) \cite{kresse_efficient_1996} within projector augmented wave
(PAW) method.\cite{blochl_projector_1994,kresse_ultrasoft_1999} The Perdew-Burke-Enzerhof (PBE) \cite{perdew_generalized_1996}
generalized gradient approximation was employed to describe exchange-correlation effects. To improve the description for localized d-electrons
in Mn$^{2+}$ ion to be strongly correlated, we added a Hubbard $U$ correction with the  fully localized limit double-counting recipe\cite{liechtenstein_density-functional_1995,dudarev_electron-energy-loss_1998}.
The effective parameter $U-J=5$ eV was used.

Additionally, as an alternative to SIESTA \cite{artacho_linear-scaling_1999}
used in Ref \cite{sanchez-portal_bonding_2002}, a similar numerical-orbital-based \cite{ozaki_variationally_2003}
DFT code OpenMX \cite{noauthor_openmx_nodate}, designed for large-scale simulation was used
for the calculations of magnetic properties. In these calculations,
core electrons are replaced with norm-conserving pseudopotential \cite{vanderbilt_soft_1990,morrison_nonlocal_1993}
and a sufficiently large numerical atomic orbital set is chosen
as Mn6.0-$s3p2d2f1$, Ba8.0-$s3p2d2$ and Bi8.0-$s3p2d2f1$ to accurately
determine the band structure. As Fig. \ref{fig:band} shows, the band structures
calculated using both codes are in good agreement.

The exchange coupling parameters are calculated using two different methods. 
We first use the standard energy-mapping approach following Ref. \cite{sanchez-portal_bonding_2002} to calculate exchange coupling constants. Due to the computation cost, and large distances between Mn sites, only the interactions up to 2nd nearest neighbors (NN) are considered. In this method, the DFT energies for the four configurations (i.e. FM and three AF) are fitted onto the following spin Hamiltonian
\begin{equation} 
H =-\sum_{\left\langle ij\right\rangle _{1}}J_{1}s_{i}s_{j}
  -\sum_{\left\langle ij\right\rangle _{1'}}J'_{1}s_{i}s_{j} 
  - \sum_{\left\langle ij\right\rangle _{2}}J_{2}s_{i}s_{j},
\label{eq:spin_ham} 
\end{equation}
where $s$ are the normalized moments, $( \mathbf{s}=\mathbf{S}/|S|$,  $|\mathbf{s}|=1$),  $J_1$,  $J_{1'}$  and $J_2$ are the exchange parameters of two inequivalent 1st NN, and 2nd NN defined in Fig. \ref{fig:structure} (a), and summation is over all different bonds.  
Note that in our definition, $J>0$ indicates a FM interaction.
This method will be applied exclusively to the data obtained using VASP and will be referred to as the ``energy difference method'' in the subsequent discussion.
 
To gain more insight,  we also employed the Green's function method \cite{katsnelson_first-principles_2000,antropov_exchange_1997}
implemented in OpenMX 3.9\cite{terasawa_efficient_2019}. In this
approach, the exchange interaction between any pair of given magnetic sites
can be directly calculated from a single magnetic state for 
any given interionic distance. 
Naturally, this approach can only be directly applied to OpenMX as it requires the Hamiltonian to be in local orbital representation.

Finally, for historical reasons, as well as in order to compare with an all-electron method, we used the Linear Augmented Wave method  Wien2K \cite{p_blaha_wien2k_2002} for the Fermi surface properties, such as the Fermi surface cuts and the plasma frequencies as a function of the Fermi energy.

\section{Doping effect on total energy\protect\label{sec:edope}}

\begin{figure}
\centerline{\includegraphics[scale=0.65]{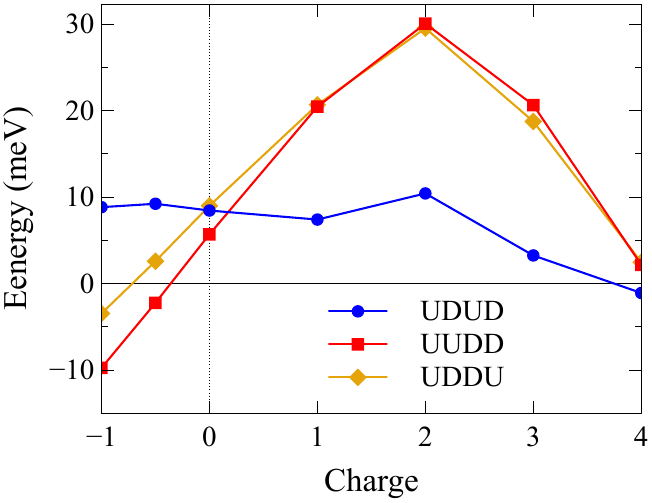}}

\caption{\protect\label{fig:e_vs_dope}Relative energy of AF configurations
compared to FM as a function of the charge doping per unit cell (4 F.U.). Negative charge represents
hole doping.}
\end{figure}

Fig. \ref{fig:e_vs_dope} shows the relative energies compared to
the FM state $uuuu$ as a function of doped charge ranging from light
hole doping up to exactly one electron per Mn (four in total)
for $udud$, $uudd$ and $uddu$ three AF orderings, where $u$ $(d)$
represents the up (down) spin moment on the Mn site and the structure
and the order of Mn are defined in Fig. \ref{fig:structure}.
The positive (negative) charge in the $x$-axis indicates electron (hole) doping. This convention is also applied to the rest of the discussion.

Generally, the relative energies for $uudd$ and $uddu$, in contrast
to that of $udud$, share a very similar pattern and both are very
sensitive to doping. In the light doping region near stoichiometry
($-1$ to 2), the trends of all three curves are rather consistent. Both
$uudd$ and $uddu$ vary monotonically and almost linearly with very steep
slopes as the energies change from around $-10$ to 30 meV within the
region while $udud$ does not show any obvious dependence and the
energy stays at roughly $9$ meV above FM.

Without doping, all three states are very close in energy and FM is
the ground state. The energy differences then begin to widen rapidly
as electrons or holes are introduced into the system. With light electron
doping, the ground state stays FM.
However, when a small fraction of holes is introduced into the system, the GS
quickly transitions into $uudd$, a very interesting type of AF,  known
as AM.
\cite{mazin_altermagnetism_2023,mazin_editorial_2022,smejkal_emerging_2022,smejkal_beyond_2022}. Indeed, as Fig. \ref{fig:structure}(d) shows, if the  $J_{1}$ bond is antiferromagnetic, there is an inversion operation connecting the two sublattices, which is a signature of a trivial AF state \cite{smolyanyuk_tool_2024} 
, while if the AF bonds are $J_{1}'$, the structure becomes AM. 

As more electrons are added to the system, a qualitative change of
behaviors can be seen in all three cases around $e=2$ (0.5 $e$ per Mn). Above this
point, the energies for all three states begin to decrease monotonically
until $e=4$, where the Fermi level reaches the gap. The system hence becomes semiconducting and
transitions into a very weakly coupled AF state in $udud$ phase.

In this gapped system all four magnetic
states (including FM) are nearly degenerate with only about 1--2
meV difference in total energy despite the large supercell. This is due to the fact that the original
exchange pathways mediated by long-range RKKY interaction in the metallic
state are no longer available. As a result the magnetic interactions
between Mn sites separated by a large distance are now instead governed
mainly by the much shorter-range superexchange mechanism, which tends
to favor AF. This $udud$ phase induced by electron doping, as previous
suggested, corresponds to the configuration that is ferromagnetic
within the same network formed by the tetrahedra units are connected
through $J_{1}$ and antiferromagnetic with adjacent network connected
by $J_{1}'$\cite {sanchez-portal_bonding_2002}.

\section{Doping effect on exchange coupling\protect\label{sec:jdope}}

\begin{figure}
\centerline{\includegraphics[scale=0.72]{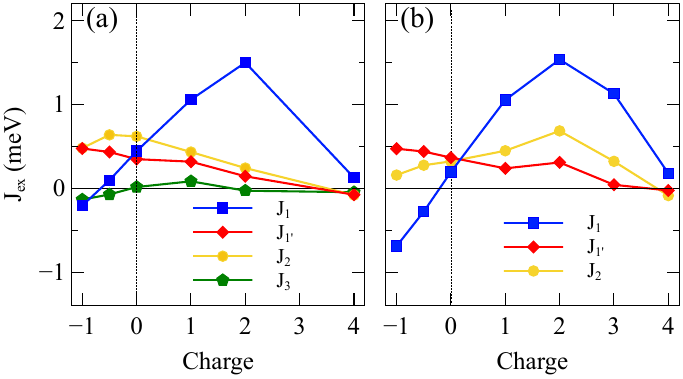}}

\caption{\protect\label{fig:jx_vs_dope}
The first few nearest exchange coupling
parameters as functions of charge doping calculated using (a) Greens function
method and (b) total energy difference. }
\end{figure}

Fig. \ref{fig:jx_vs_dope} shows the exchange coupling parameters
as functions of doping for the first few nearest NN using both (a)
GF function method and (b) total energy difference method. For the
latter, due to the computation limitation by the system size, only
up to $J_{2}$ are calculated.

Generally, some differences between the two methods are expected,
as in the case of the energy difference method, the further NN interactions
which could still contribute in the RKKY-driven regime are combined
into the nearest three exchange parameters, while GF is a perturbation
method on a single determinant which allows the calculation between
any given pair of any distance.  
Also, these two methods differ in that the perturbative Green function method corresponds to small deviations of collinearity and the total energy method assumes full spin flips; the two may physically differ if the magnetic Hamiltonian deviates from the Heisenberg form

Despite the difference, several key features are rather consistently
predicted by both methods and offer useful insights
into understanding the effect of doping. Most importantly, both suggest
that $J_{1}$ is the most susceptible component to doping and 
its sign and strength vary greatly depending on the type of doping
while $J_{1}'$ and $J_{2}$ change more gradually. Other qualitative
behaviors such as the crossover between the $J_{1}'$ and $J_{2}$
as well as the sign change of $J_{1}$ in the hole doping region are
also captured.

Both $uudd$ and $uddu$ respond to doping in a very similar way
suggest that they share the same dominant mechanism that is very different
from that of $udud$. Indeed, from both Figs \ref{fig:jx_vs_dope}
(a) and (b), one can see that the qualitative behaviors of $uudd$
and $uddu$ closely resemble that of $J_{1}$ which connect opposite
spins in both states indicating that $J_{1}$ is the main driving
mechanism of the dependency.

Although other major components such as $J_{1}'$ and $J_{2}$ are
also significant in terms of the strength, the effects from $J_{1}'$ and farther NN interaction
$J_{2}$ roughly cancel out. This cancellation is mainly due to the
fact that, while the strength of $J_{2}$ is about twice bigger than
$J_{1}'$ but, for any given Mn site, the number of its 2nd NN is
only of half its first NN  connected through $J_{1}'$.
This explains why $uudd$ and $uddu$ nearly degenerate
in electron doping cases.

However, in the hole doping case, this cancellation no longer holds,
as $J_{1}'$ continues to increase with more hole carriers introduced
into the system, while $J_{2}$ reaches the peak and begin to decline.
As a result, in Fig. \ref{fig:e_vs_dope}, one can
see the energy gap between $uudd$ and $uddu$ begins to widen and
$uudd$ becomes the ground state. It is interesting to note that,
despite some discrepancy in the qualitative behavior, both Figs \ref{fig:jx_vs_dope}
(a) and (b) exhibit the same cancellation in charge doping region
and the cancellation is removed due to hole doping.

On the other hand, in $udud$, the energy term that involves $J_{1}$,
the most sensitive component to doping, has the same sign as that
of the FM ordering, therefore the doping dependence of the energy
difference between the two, is mainly influenced by other exchange
paths, such as less sensitive $J_{1}'$ and $J_{2}$ . 

\begin{figure}
\centerline{\includegraphics[scale=0.78]{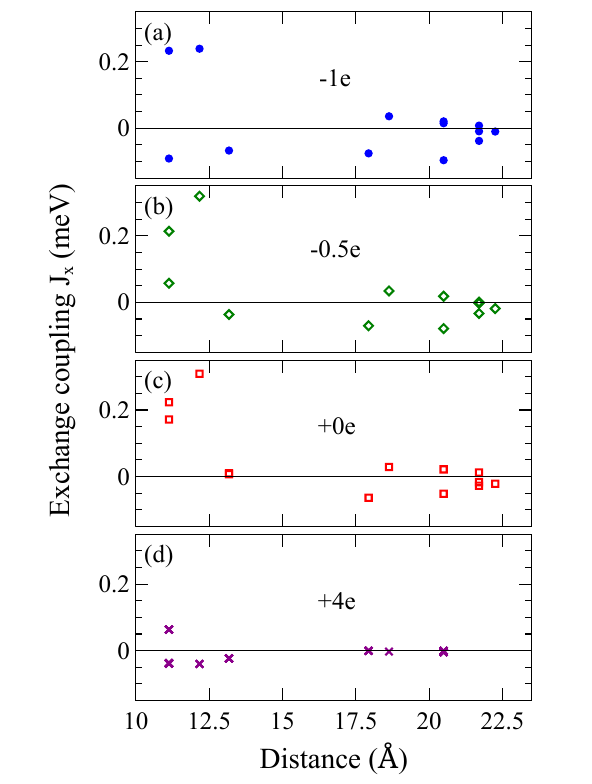}}
\caption{\protect\label{fig:j_vs_d}Exchange coupling constants as functions
of distance for different amounts of doping calculated using Green's
function method.}
\end{figure}

Fig. \ref{fig:j_vs_d} shows the exchange coupling parameters as functions
of distance for Bi$_{14}$MnBi$_{11}$ with different amounts of doping,
calculated using the Green's function method. The exchange coupling parameters
in all three metallic states ($0$, $0.5$ and $1$) possess long-range (there are still significant contributions beyond
20 \AA)
and sign alternating behavior which are both signatures of the RKKY interaction where the localized $3d$ Mn moments are coupled through conduction electrons.
This is likely the reason for the unusually high $T_{N}$
given such large spacing between the magnetic Mn sites. 

For $e=+4$, the system in a semiconducting state as mentioned earlier,
likely governed by super-exchange, has only much weaker and shorter
range interactions that favor AF, as shown in Fig. \ref{fig:j_vs_d}(d)
all $J_{1-3}$ are much smaller, and essentially negligible beyond
$J_{3}$.

However, it is important to note that, while there are still significant
interactions beyond 3NN ($\geq 14$ \AA)
in the metallic states, most of the farther neighbors are rather insensitive
to doping as shown in Figs. \ref{fig:j_vs_d}(a)-(c). It is therefore reasonable
to assume that the qualitative behaviors due to doping are mainly
dictated by $J_{1}$, $J_{1}'$ and $J_{2}$. 

\begin{figure}
\centerline{\includegraphics[scale=0.70]{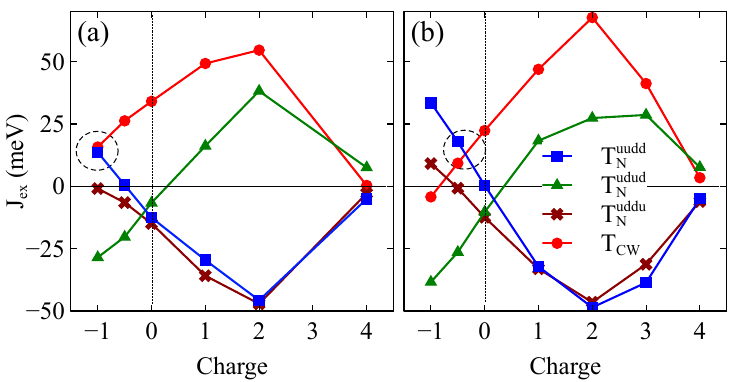}}
\caption{\protect\label{fig:Tcw}
Mean-field $T_{CW}$ and $T_N$ as functions of doping using (a) total energy (VASP) and (b) Green's function method (OpenMX)
}
\end{figure}

\subsection*{
$T_{CW}$ and $T_N$
}


\begin{table}
    \renewcommand*{\arraystretch}{1.3}
    \centering
    \begin{tabular}{|>{\centering\arraybackslash}p{0.15\linewidth}|>{\centering\arraybackslash}p{0.12\linewidth}|>{\centering\arraybackslash}p{0.12\linewidth}|>{\centering\arraybackslash}p{0.12\linewidth}|} \hline 
          & $J_1$& $J_{1'}$&$J_2$\\ \hline 
         CW& $+$& $+$&$+$\\ \hline  
         FM&  $+$&  $+$&  $+$\\ \hline  
         $udud$&  $+$&  $-$&  $-$\\ \hline  
         $uudd$&  $-$&  $+$&  $-$\\ \hline  
         $uddu$&  $-$&  $-$&  $+$\\ \hline 
    \end{tabular}
    \caption{The sign ($\sigma=s_is_j$) indicates the alignment between the magnetic moments connected by a given $J$ for each magnetic state.}
    \label{tab:j_sign}
\end{table}
All Mn sites are equivalent, and each Mn is connected to eight first NN: four through $J_1$ and four through $J_{1'}$, as well as two second NN through $J_2$. The critical temperatures $T_c$ (c=CW and N for Curie-Weiss and N\`{e}el temperature respectively) in the mean-field approximation are directly related to the strength of exchange interactions $J_i$ and can be expressed in the following simple form
\begin{equation} 
T_{c}=\frac{2}{3k_B}(4J_1\sigma_1+4J_{1'}\sigma_{1'}+2J_2\sigma_{2}),
\label{eq:Tcw} 
\end{equation}
where $k_B$ is the Boltzmann constant and $\sigma=\boldsymbol{s_is_j}$ are the signs, as summarized in TABLE \ref{tab:j_sign}, indicate the alignment between the magnetic moments connected by the given $J_i$ in each magnetic state.
%
Figure \ref{fig:Tcw} shows the mean-field $T_{N}$ for the three AF states and $T_{CW}$ 
estimated from the exchange parameters obtained
using (a) Green's function method and (b) total energy difference method as presented in Fig. \ref{fig:jx_vs_dope}.
It is worth noting that the Curie temperature $T_C$ for $uuuu$ has an identical expression as  $T_{CW}$ in the mean-field approximation.
Despite some small deviations in $J_2$, both methods agree reasonably well suggesting that the Mn moments are fairly localized.
Generally, the state with the highest critical temperature ($T_N$ or $T_C$) corresponds to the most stable and energetically favorable state.   
This becomes clear when compared with Figs. \ref{fig:e_vs_dope} and \ref{fig:jx_vs_dope}.

Light electron doping stabilizes the FM state leading to a positive and increasing $T_{CW}$. At $e=4$ all temperatures become very small as a result of weak couplings in semiconducting states, with $udud$ being the ground state and $T_N^{udud}$ having the largest value.
On the other hand, light hole doping has the opposite effect promoting $uudd$ while destabilizing the FM state.
Given the close resemblance between $uudd$ and $udud$ shown in Fig. \ref{fig:e_vs_dope}, it is not surprising that their  $T_N$ values also have the same dependence on doping, despite $uudd$ being more stable.

Most importantly, near the hole doping concentration where $uudd$ (AM) is predicted to be the ground state 
(i.e. $e=1.0$ in Fig. (a) and $e=0.4$ in Fig. (b)),  two curves $T_{CW}$ and $T_N^{uudd}$ intersect at around $T=15$ K. 
This naturally implies that both temperatures have the same sign and are very close in magnitude ($\sim$15 K) in the proximity of the intersection, as indicated by the dashed circles  in both Figs. \ref{fig:Tcw} (a) and (b).
This is indeed supported by the experimental values showing $T_N=15$ K and $T_{CW}=17$ K \cite{kuromoto_structure_1992}.
This finding is rather intriguing because a positive $T_{CW}$ suggests a FM state which contradicts a positive $T_N$ that indicates an AF state.
As introduced earlier, this phenomenon is common in the antiferromagnets consisting of strongly coupled FM planes with a weak AF interaction between them.  

Although, at first glance, the highly symmetric crystal structure of Ba$_{14}$MnBi$_{11}$  appears three-dimensional and does not allow such type of planer coupling, the magnetic moments, instead of planar, can in fact be visualized as two interpenetrating 3D networks \cite{sanchez-portal_bonding_2002}.

The discrepancy can then be easily explained if the Mn moments within the same network, coupled through $J_{1'}$, strongly favor the FM ordering, meanwhile a weak AF interaction, coupled through $J_{1}$, exists between networks. This explanation is consistent with our data shown in Fig. \ref{fig:jx_vs_dope} further supporting the argument made in previous work \cite{sanchez-portal_bonding_2002}.
However, it is important to note that, the definition of Mn network here is different from that in Ref. \cite{sanchez-portal_bonding_2002}, where they consider the Mn tetrahedra complexes connected by $J_{1}$ to belong to the same network.


\section{Exchange pathway and band characters\protect\label{sec:jpath}}

Due to the large distances that separate the Mn atoms, the behaviors
of the exchange couplings are expected to be heavily influenced by
the detailed electronic structure of Bi-$p$ bands as they are the
major components near the Fermi level \cite{sanchez-portal_bonding_2002}.
To gain more insight into the underlying mechanism, it is helpful to
first look into the exchange pathway and identify the band characters
associated with it. 

Fig. \ref{fig:structure}(a) shows the definitions of the three dominant
exchange parameters $J_{1}$, $J_{1'}$ and $J_{2}$ along with the
relavant surrouding elements displayed. One can see a clear distinction
between the three. For $J_{1}$, the path directly goes through two
Bi3 sites while $J_{1'}$ does not overlap with any sites directly.
However, its exchange pathway likely involves two Bi4 atoms near
the path that form a plane with the two Mn sites connected by $J_{1'}$.
The Next NN $J_{2}$ has a direct overlap with the Bi1 at the midpoint
which is likely to involve the entire trimer (Bi1 and Bi2). As the Bi band
consists of four different inequivalent Bi-sites as introduced in Section
\ref{sec:structure}, one can immediately notice that each pathway
can be associated with at least one distinct type of Bi character
that is directly involved.

%
%
\begin{figure}
\centerline{\includegraphics[scale=0.63]{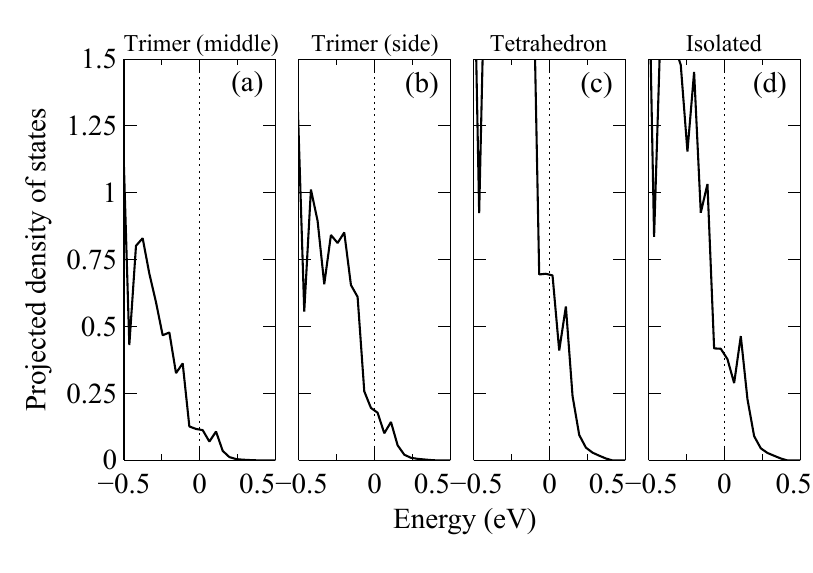}}
\caption{\protect\label{fig:pdos}Projected density of state of four inequivalent
Bi sites near the Fermi level.}
\end{figure}

Fig.\ref{fig:pdos} shows the projected density of states (pDOS) for all four inequivalent
Bi sites near the Fermi level and valence band maximum (VBM). Generally, the Bi sites 
from the tetrahedral units have the largest contribution around the
Fermi level and host the biggest portion of the hole concentration
(near VBM). This explains why $J_{1}$ is the most susceptible parameter
to band filling. In addition, there is also a significant amount of
isolated Bi (Bi2). In comparison, the
trimer characters are less significant near the Fermi level. This
configuration of orbital characters is not uncommon. A similar arrangement
is also seen in other Zintel 14-11-1 compounds, as the bandstructures
closely resemble one another \cite{liu_finding_2021}.

It is worth mentioning that, our estimation of DOS for FM state at $E_F$ is approximately 7.0 states/eV$\cdot$formula which is very close to the earlier theoretical value (7.8 states/eV$\cdot$formula) \cite{sanchez-portal_bonding_2002}. 
The large difference compared to the experimental value ($\sim44$ states/eV$\cdot$formula) \cite{siemens_specific_1992} has been attributed to the effect of hyperfine splitting not being considered
However, the mechanism behind the unexpectedly large observed DOS remains unclear.

%
%
\begin{figure}[h]
\centerline{
\includegraphics[scale=0.31]{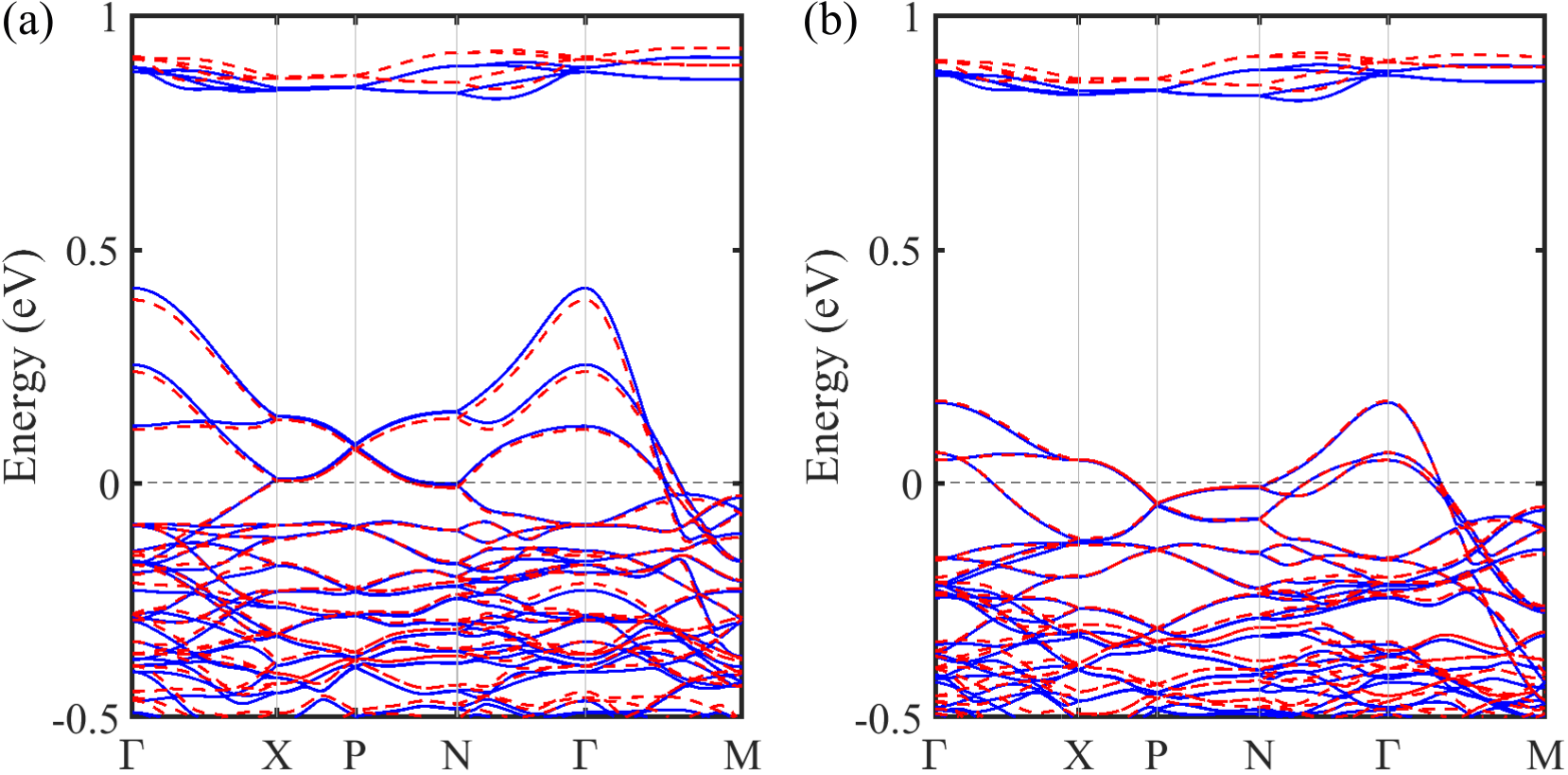}
}
\caption{\protect\label{fig:band} 
The bandstructure of Ba$_{14}$MnBi$_{11}$ 
in FM states for (a) majority and (b) minority spins using VASP (red) and OpenMX (blue).
}
\end{figure}
Fig. \ref{fig:band} shows the bandstructure near the VBM for Ba$_{14}$MnBi$_{11}$
in FM state for both (a) majority and (b) minority spins {[}calculated
using VASP (red) and OpenMX (black) as an additional check. The two
methods yield remarkable agreement{]}. 

Similar to the previous findings \cite{liu_finding_2021,sanchez-portal_bonding_2002},
near the Fermi level a few rather dispersive light bands
emerge from the flatter valence at $\Gamma$ point both seen in (a)
and (b), indicates a p-type conductor. As shown in the pDOS analysis Fig. \ref{fig:pdos},
these bands near VBM are mainly made up of the Bi characters from
the tetrahedra unit with significant contribution from the isolated
Bi sites. The relative high discursiveness of the hole pockets around
$\Gamma$ point suggests a more complicated hybridization network
most likely between the two dominant characters.


%
%
\section{Electronic and transport properties}
\begin{figure}[h]
\includegraphics[scale=0.6]{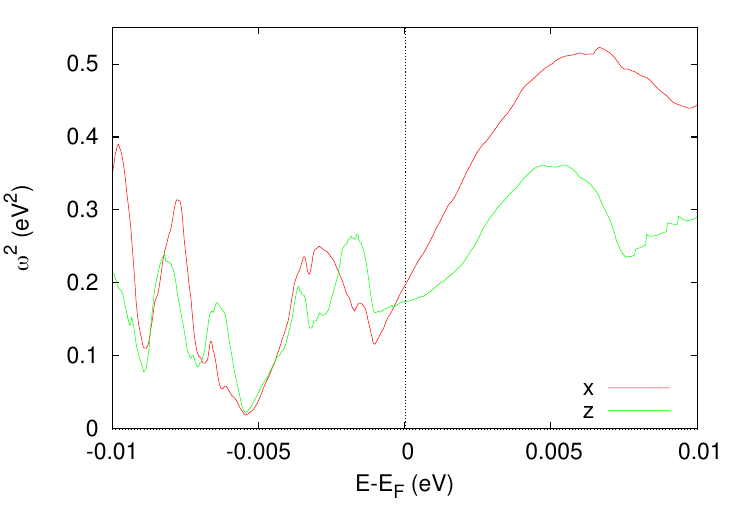}
\caption{\protect\label{fig:am_wp2} 
The plasma frequency as a function of the Fermi level calculated along $x$ and $z$ directions.
}
\end{figure}

In the transport measurement, the temperature dependence of RRR indicates metallic behavior, as resistivity decreases with the reducing temperature, despite the high resistivity observed.
In the original work, this high resistivity is attributed to the polycrystal sample \cite{kuromoto_structure_1992}. 
However, our transport calculations suggest that the compound inherently has poor conductivity regardless of sample quality.
Figure \ref{fig:am_wp2} shows the plasma frequency as a function of Fermi energy. The values calculated along both $x$ (red) and $z$ (green) directions are generally smaller than 0.5 eV$^2$, which is at least an order of magnitude smaller than in typical metals.
In the Drude approximation, the static conductivity $\sigma_0$ is determined by 
\begin{equation} 
    \sigma_0=\omega^2_p {\tau}/{4\pi}
\end{equation} 
where $\tau$ and $\omega_p$  are the relaxation time and the plasma frequency, respectively. Such a small
$\omega_p^2$ inherently leads to a small conductivity, although a short relaxation time can also contribute.

To better understand magnetism in \BNB,
it is also important to highlight the intricate connection between the magnetic and electronic structures. 
As explained in detail in Ref. \cite{sanchez-portal_bonding_2002}, a comparison with the electronic structure of a similar compound Ca$_{14}$GaAs$_{11}$ (where Ga is identified to be trivalent) \cite{gallup_bonding_1992}, reveals that if Mn were in a trivalent state, the system would have been semiconducting \cite{gallup_bonding_1992}, which contradicts the transport measurements \cite{kuromoto_structure_1992}.
To be consistent with a metallic band structure,   Mn should instead be in the divalent state (Mn$^{2+}$) $d^5$ configuration, which is consistent with the DFT calculations. The deviation from the experimentally observed net magnetization of $\sim4$ $\mu_{B}$ is attributed to the opposite polarization of the Bi atoms that form tetrahedral units with Mn. Therefore, the entire MnBi$_4^{-}$  complex should be treated as one magnetic unit in order to explain the discrepancy.
This conclusion is further supported by our Voronoi charge analysis, which shows a clear $d^5$ Mn state ($\sim4.8\mu_B$) and the opposite polarization on Bi sites ($\sim0.03-0.05$ $\mu_B$/Bi), consistent with the conclusions of Ref.\cite{sanchez-portal_bonding_2002}.

%
%
Intuitively, the FM order obtained from DFT is related to
the metallicity of the stoichiometric compound; indeed, adding one additional electron
per F.U to the system opens a semiconducting gap and the ground states 
transitions into a weakly coupled AF state that corresponds to the
state proposed in Ref. \cite{sanchez-portal_bonding_2002}. However, this does not account
for the observed metallicity (albeit this observation may be questioned) and the 
weak interactions hardly account for the $T_{N}$ observed.

%
%
On the other hand, we found that the magnetic states are susceptible
to band filling particularly near stoichiometry. So, through a small
amount of hole doping, we were able to obtain another type of AF state
known as altermagmetic, while preserving the metallic conductivity. We therefore believe
that the discrepancy in the magnetic state between DFT calculations and
experimental findings stems from nonstoichiometry and intrinsic small hole doping and the AF state observed in the experiment is in fact altermagnetic. 

\begin{figure}[h]
\includegraphics[scale=2.0]{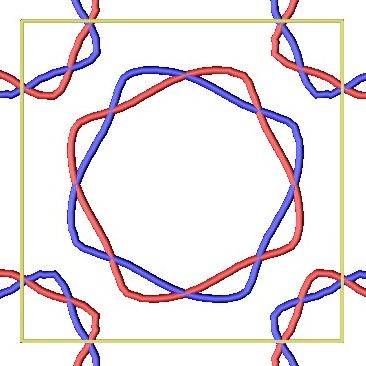}
\caption{\protect\label{fig:band_split_am} 
The cross-section of Fermi surfaces at $k_z=0$ in the altermagnetic state. The red and blue contours represent up and down spins, respectively.
}
\end{figure}

Figure \ref{fig:band_split_am} shows the cross-section of Fermi surfaces at $k_z=0$ of the AM state. The red (blue) color represents the up (down) spin. The band splits in a compensated magnetic order is the signature of altermagnetism.
 
\section{Conclusions and considerations regarding altermagetism\protect\label{sec:conclusion}}

In this study, we conducted an extensive and comprehensive examination
of the magnetic properties of Ba$_{14}$MnBi$_{11}$ using several different
Density Functional Theory (DFT) codes.
We have successfully established a connection between the electronic and magnetic structures,
and were able to explain some of the unresolved issues.
Our investigation revealed a significant sensitivity of the first
few nearest exchange coupling interactions to band filling. With either
electron or hole doping, we observed the emergence of distinct antiferromagnetic
(AF) states. 
By adding exactly one electron per F.U., which fills
the valence hole pocket, the system transitions into an AF state
where the moments are ferromagnetically coupled through $J_1$ and antiferromagnetically coupled through  $J_{1'}$.
Conversely, the introduction of
light hole doping results in a unique type of AF state known as altermagnetism, 
which simultaneously also preserves the metallic state observed in the 
experiment.  
In this case, the Ne\'{e}l and Curie-Weiss temperatures obtained from mean-field 
approximation are both in excellent agreement with the experimental 
data.
Based on our analysis, 
the AF observed state is expected to be altermagnetic and
the discrepancy in the magnetic
ground state between the DFT calculations and experimental observation
is likely due to deviations from the stoichiometry such as Ba vacancies. 

Importantly, the predicted altermagnetic state has a number of interesting properties suggesting that if better quality samples can be manufactured they can provide considerable advantage over existing altermagnetic candidates. First, it belongs to the $d-$wave class, per classification of Ref. \cite{smejkal_beyond_2022}, which reduces the number of vertical nodal lines (compared to the two other convincingly established semiconducting/metallic altermagnets, MnTe and CrSb). Most importantly, despite its relatively high tetragonal symmetry, it has non-zero anomalous transport for all possible orientations of the N\'eel vector. Specifically, in the case of the easy-axis anisotropy, $\mathbf{M}||c$, anomalous conductivity $\sigma_{xz}=\sigma_{xz}\neq 0$ is allowed, and for the easy plane anisotropy,  $\mathbf{M}\perp c$, all three components are non-zero, $\sigma_{xz}=\sigma_{xz}\neq \sigma_{xy}\neq 0$. This property is quite unique among altermagnetic candidates. The low N\'eel temperature of 15 K precludes practical application, but this material remains an interesting candidate to study altermagnetic physics.

\begin{acknowledgments}
This work was supported by the Army Research Office under Cooperative Agreement Number W911NF-22-2-0173.
\end{acknowledgments}



\bibliography{zintl_altermag}

\begin{thebibliography}{27}%
\makeatletter
\providecommand \@ifxundefined [1]{%
 \@ifx{#1\undefined}
}%
\providecommand \@ifnum [1]{%
 \ifnum #1\expandafter \@firstoftwo
 \else \expandafter \@secondoftwo
 \fi
}%
\providecommand \@ifx [1]{%
 \ifx #1\expandafter \@firstoftwo
 \else \expandafter \@secondoftwo
 \fi
}%
\providecommand \natexlab [1]{#1}%
\providecommand \enquote  [1]{``#1''}%
\providecommand \bibnamefont  [1]{#1}%
\providecommand \bibfnamefont [1]{#1}%
\providecommand \citenamefont [1]{#1}%
\providecommand \href@noop [0]{\@secondoftwo}%
\providecommand \href [0]{\begingroup \@sanitize@url \@href}%
\providecommand \@href[1]{\@@startlink{#1}\@@href}%
\providecommand \@@href[1]{\endgroup#1\@@endlink}%
\providecommand \@sanitize@url [0]{\catcode `\\12\catcode `\$12\catcode
  `\&12\catcode `\#12\catcode `\^12\catcode `\_12\catcode `\%12\relax}%
\providecommand \@@startlink[1]{}%
\providecommand \@@endlink[0]{}%
\providecommand \url  [0]{\begingroup\@sanitize@url \@url }%
\providecommand \@url [1]{\endgroup\@href {#1}{\urlprefix }}%
\providecommand \urlprefix  [0]{URL }%
\providecommand \Eprint [0]{\href }%
\providecommand \doibase [0]{https://doi.org/}%
\providecommand \selectlanguage [0]{\@gobble}%
\providecommand \bibinfo  [0]{\@secondoftwo}%
\providecommand \bibfield  [0]{\@secondoftwo}%
\providecommand \translation [1]{[#1]}%
\providecommand \BibitemOpen [0]{}%
\providecommand \bibitemStop [0]{}%
\providecommand \bibitemNoStop [0]{.\EOS\space}%
\providecommand \EOS [0]{\spacefactor3000\relax}%
\providecommand \BibitemShut  [1]{\csname bibitem#1\endcsname}%
\let\auto@bib@innerbib\@empty
\bibitem [{\citenamefont {S{\'a}nchez-Portal}\ \emph
  {et~al.}(2002)\citenamefont {S{\'a}nchez-Portal}, \citenamefont {Martin},
  \citenamefont {Kauzlarich},\ and\ \citenamefont
  {Pickett}}]{sanchez-portal_bonding_2002}%
  \BibitemOpen
  \bibfield  {author} {\bibinfo {author} {\bibfnamefont {D.}~\bibnamefont
  {S{\'a}nchez-Portal}}, \bibinfo {author} {\bibfnamefont {R.~M.}\ \bibnamefont
  {Martin}}, \bibinfo {author} {\bibfnamefont {S.~M.}\ \bibnamefont
  {Kauzlarich}},\ and\ \bibinfo {author} {\bibfnamefont {W.~E.}\ \bibnamefont
  {Pickett}},\ }\bibfield  {title} {\bibinfo {title} {Bonding, moment
  formation, and magnetic interactions in {Ca}$_{14}${MnBi}$_{11}$ and
  {Ba}$_{14}${MnBi}$_{11}$},\ }\href@noop {} {\bibfield  {journal} {\bibinfo
  {journal} {Phys. Rev. B}\ }\textbf {\bibinfo {volume} {65}},\ \bibinfo
  {pages} {144414} (\bibinfo {year} {2002})}\BibitemShut {NoStop}%
\bibitem [{\citenamefont {Webb}\ \emph {et~al.}(1991)\citenamefont {Webb},
  \citenamefont {Kuromoto},\ and\ \citenamefont {Kauzlarich}}]{webb_new_1991}%
  \BibitemOpen
  \bibfield  {author} {\bibinfo {author} {\bibfnamefont {D.~J.}\ \bibnamefont
  {Webb}}, \bibinfo {author} {\bibfnamefont {T.~Y.}\ \bibnamefont {Kuromoto}},\
  and\ \bibinfo {author} {\bibfnamefont {S.~M.}\ \bibnamefont {Kauzlarich}},\
  }\bibfield  {title} {\bibinfo {title} {New ternary magnets ({Ca}, {Sr},
  {Ba}){14MnBi11}},\ }\href {https://doi.org/10.1016/0304-8853(91)90429-E}
  {\bibfield  {journal} {\bibinfo  {journal} {Journal of Magnetism and Magnetic
  Materials}\ }\textbf {\bibinfo {volume} {98}},\ \bibinfo {pages} {71}
  (\bibinfo {year} {1991})}\BibitemShut {NoStop}%
\bibitem [{\citenamefont {Kuromoto}\ \emph {et~al.}(1992)\citenamefont
  {Kuromoto}, \citenamefont {Kauzlarich},\ and\ \citenamefont
  {Webb}}]{kuromoto_structure_1992}%
  \BibitemOpen
  \bibfield  {author} {\bibinfo {author} {\bibfnamefont {T.~Y.}\ \bibnamefont
  {Kuromoto}}, \bibinfo {author} {\bibfnamefont {S.~M.}\ \bibnamefont
  {Kauzlarich}},\ and\ \bibinfo {author} {\bibfnamefont {D.~J.}\ \bibnamefont
  {Webb}},\ }\bibfield  {title} {\bibinfo {title} {Structure and properties of
  the transition-metal zintl compounds: {A14MnBi11} ({A} = {Ca}, {Sr}, {Ba})},\
  }\href {https://doi.org/10.1021/cm00020a036} {\bibfield  {journal} {\bibinfo
  {journal} {Chem. Mater.}\ }\textbf {\bibinfo {volume} {4}},\ \bibinfo {pages}
  {435} (\bibinfo {year} {1992})}\BibitemShut {NoStop}%
\bibitem [{\citenamefont {Kauzlarich}\ \emph {et~al.}(2004)\citenamefont
  {Kauzlarich}, \citenamefont {Payne},\ and\ \citenamefont
  {Webb}}]{miller_magnetism_2004}%
  \BibitemOpen
  \bibfield  {author} {\bibinfo {author} {\bibfnamefont {S.~M.}\ \bibnamefont
  {Kauzlarich}}, \bibinfo {author} {\bibfnamefont {A.~C.}\ \bibnamefont
  {Payne}},\ and\ \bibinfo {author} {\bibfnamefont {D.~J.}\ \bibnamefont
  {Webb}},\ }\bibfield  {title} {\bibinfo {title} {Magnetism and
  {Magnetotransport} {Properties} of {Transition} {Metal} {Zintl} {Isotypes}},\
  }in\ \href {https://doi.org/10.1002/9783527620548.ch2b} {\emph {\bibinfo
  {booktitle} {Magnetism: {Molecules} to {Materials}}}},\ \bibinfo {editor}
  {edited by\ \bibinfo {editor} {\bibfnamefont {J.~S.}\ \bibnamefont {Miller}}\
  and\ \bibinfo {editor} {\bibfnamefont {M.}~\bibnamefont {Drillon}}}\
  (\bibinfo  {publisher} {Wiley},\ \bibinfo {year} {2004})\ \bibinfo {edition}
  {1st}\ ed.,\ pp.\ \bibinfo {pages} {37--62}\BibitemShut {NoStop}%
\bibitem [{\citenamefont {Mazin}(2023)}]{mazin_altermagnetism_2023}%
  \BibitemOpen
  \bibfield  {author} {\bibinfo {author} {\bibfnamefont {I.~I.}\ \bibnamefont
  {Mazin}},\ }\bibfield  {title} {\bibinfo {title} {Altermagnetism in {MnTe}:
  {Origin}, predicted manifestations, and routes to detwinning},\ }\href
  {https://doi.org/10.1103/PhysRevB.107.L100418} {\bibfield  {journal}
  {\bibinfo  {journal} {Phys. Rev. B}\ }\textbf {\bibinfo {volume} {107}},\
  \bibinfo {pages} {L100418} (\bibinfo {year} {2023})}\BibitemShut {NoStop}%
\bibitem [{\citenamefont {Mazin}\ and\ \citenamefont {{The PRX
  Editors}}(2022)}]{mazin_editorial_2022}%
  \BibitemOpen
  \bibfield  {author} {\bibinfo {author} {\bibfnamefont {I.}~\bibnamefont
  {Mazin}}\ and\ \bibinfo {author} {\bibnamefont {{The PRX Editors}}},\
  }\bibfield  {title} {\bibinfo {title} {Editorial:
  {Altermagnetism}{\textemdash}{A} {New} {Punch} {Line} of {Fundamental}
  {Magnetism}},\ }\href {https://doi.org/10.1103/PhysRevX.12.040002} {\bibfield
   {journal} {\bibinfo  {journal} {Phys. Rev. X}\ }\textbf {\bibinfo {volume}
  {12}},\ \bibinfo {pages} {040002} (\bibinfo {year} {2022})}\BibitemShut
  {NoStop}%
\bibitem [{\citenamefont {{\v S}mejkal}\ \emph
  {et~al.}(2022{\natexlab{a}})\citenamefont {{\v S}mejkal}, \citenamefont
  {Sinova},\ and\ \citenamefont {Jungwirth}}]{smejkal_emerging_2022}%
  \BibitemOpen
  \bibfield  {author} {\bibinfo {author} {\bibfnamefont {L.}~\bibnamefont {{\v
  S}mejkal}}, \bibinfo {author} {\bibfnamefont {J.}~\bibnamefont {Sinova}},\
  and\ \bibinfo {author} {\bibfnamefont {T.}~\bibnamefont {Jungwirth}},\
  }\bibfield  {title} {\bibinfo {title} {Emerging {Research} {Landscape} of
  {Altermagnetism}},\ }\href {https://doi.org/10.1103/PhysRevX.12.040501}
  {\bibfield  {journal} {\bibinfo  {journal} {Phys. Rev. X}\ }\textbf {\bibinfo
  {volume} {12}},\ \bibinfo {pages} {040501} (\bibinfo {year}
  {2022}{\natexlab{a}})}\BibitemShut {NoStop}%
\bibitem [{\citenamefont {{\v S}mejkal}\ \emph
  {et~al.}(2022{\natexlab{b}})\citenamefont {{\v S}mejkal}, \citenamefont
  {Sinova},\ and\ \citenamefont {Jungwirth}}]{smejkal_beyond_2022}%
  \BibitemOpen
  \bibfield  {author} {\bibinfo {author} {\bibfnamefont {L.}~\bibnamefont {{\v
  S}mejkal}}, \bibinfo {author} {\bibfnamefont {J.}~\bibnamefont {Sinova}},\
  and\ \bibinfo {author} {\bibfnamefont {T.}~\bibnamefont {Jungwirth}},\
  }\bibfield  {title} {\bibinfo {title} {Beyond {Conventional} {Ferromagnetism}
  and {Antiferromagnetism}: {A} {Phase} with {Nonrelativistic} {Spin} and
  {Crystal} {Rotation} {Symmetry}},\ }\href
  {https://doi.org/10.1103/PhysRevX.12.031042} {\bibfield  {journal} {\bibinfo
  {journal} {Phys. Rev. X}\ }\textbf {\bibinfo {volume} {12}},\ \bibinfo
  {pages} {031042} (\bibinfo {year} {2022}{\natexlab{b}})}\BibitemShut
  {NoStop}%
\bibitem [{\citenamefont {Kresse}\ and\ \citenamefont
  {Furthm{\"u}ller}(1996)}]{kresse_efficient_1996}%
  \BibitemOpen
  \bibfield  {author} {\bibinfo {author} {\bibfnamefont {G.}~\bibnamefont
  {Kresse}}\ and\ \bibinfo {author} {\bibfnamefont {J.}~\bibnamefont
  {Furthm{\"u}ller}},\ }\bibfield  {title} {\bibinfo {title} {Efficient
  iterative schemes for ab initio total-energy calculations using a plane-wave
  basis set},\ }\href {https://doi.org/10.1103/PhysRevB.54.11169} {\bibfield
  {journal} {\bibinfo  {journal} {Phys. Rev. B}\ }\textbf {\bibinfo {volume}
  {54}},\ \bibinfo {pages} {11169} (\bibinfo {year} {1996})},\ \bibinfo {note}
  {publisher: American Physical Society}\BibitemShut {NoStop}%
\bibitem [{\citenamefont {Bl{\"o}chl}(1994)}]{blochl_projector_1994}%
  \BibitemOpen
  \bibfield  {author} {\bibinfo {author} {\bibfnamefont {P.~E.}\ \bibnamefont
  {Bl{\"o}chl}},\ }\bibfield  {title} {\bibinfo {title} {Projector
  augmented-wave method},\ }\href {https://doi.org/10.1103/PhysRevB.50.17953}
  {\bibfield  {journal} {\bibinfo  {journal} {Phys. Rev. B}\ }\textbf {\bibinfo
  {volume} {50}},\ \bibinfo {pages} {17953} (\bibinfo {year} {1994})},\
  \bibinfo {note} {publisher: American Physical Society}\BibitemShut {NoStop}%
\bibitem [{\citenamefont {Kresse}\ and\ \citenamefont
  {Joubert}(1999)}]{kresse_ultrasoft_1999}%
  \BibitemOpen
  \bibfield  {author} {\bibinfo {author} {\bibfnamefont {G.}~\bibnamefont
  {Kresse}}\ and\ \bibinfo {author} {\bibfnamefont {D.}~\bibnamefont
  {Joubert}},\ }\bibfield  {title} {\bibinfo {title} {From ultrasoft
  pseudopotentials to the projector augmented-wave method},\ }\href
  {https://doi.org/10.1103/PhysRevB.59.1758} {\bibfield  {journal} {\bibinfo
  {journal} {Phys. Rev. B}\ }\textbf {\bibinfo {volume} {59}},\ \bibinfo
  {pages} {1758} (\bibinfo {year} {1999})},\ \bibinfo {note} {publisher:
  American Physical Society}\BibitemShut {NoStop}%
\bibitem [{\citenamefont {Perdew}\ \emph {et~al.}(1996)\citenamefont {Perdew},
  \citenamefont {Burke},\ and\ \citenamefont
  {Ernzerhof}}]{perdew_generalized_1996}%
  \BibitemOpen
  \bibfield  {author} {\bibinfo {author} {\bibfnamefont {J.~P.}\ \bibnamefont
  {Perdew}}, \bibinfo {author} {\bibfnamefont {K.}~\bibnamefont {Burke}},\ and\
  \bibinfo {author} {\bibfnamefont {M.}~\bibnamefont {Ernzerhof}},\ }\bibfield
  {title} {\bibinfo {title} {Generalized {Gradient} {Approximation} {Made}
  {Simple}},\ }\href {https://doi.org/10.1103/PhysRevLett.77.3865} {\bibfield
  {journal} {\bibinfo  {journal} {Phys. Rev. Lett.}\ }\textbf {\bibinfo
  {volume} {77}},\ \bibinfo {pages} {3865} (\bibinfo {year} {1996})},\ \bibinfo
  {note} {publisher: American Physical Society}\BibitemShut {NoStop}%
\bibitem [{\citenamefont {Liechtenstein}\ \emph {et~al.}(1995)\citenamefont
  {Liechtenstein}, \citenamefont {Anisimov},\ and\ \citenamefont
  {Zaanen}}]{liechtenstein_density-functional_1995}%
  \BibitemOpen
  \bibfield  {author} {\bibinfo {author} {\bibfnamefont {A.~I.}\ \bibnamefont
  {Liechtenstein}}, \bibinfo {author} {\bibfnamefont {V.~I.}\ \bibnamefont
  {Anisimov}},\ and\ \bibinfo {author} {\bibfnamefont {J.}~\bibnamefont
  {Zaanen}},\ }\bibfield  {title} {\bibinfo {title} {Density-functional theory
  and strong interactions: {Orbital} ordering in {Mott}-{Hubbard} insulators},\
  }\href {https://doi.org/10.1103/PhysRevB.52.R5467} {\bibfield  {journal}
  {\bibinfo  {journal} {Phys. Rev. B}\ }\textbf {\bibinfo {volume} {52}},\
  \bibinfo {pages} {R5467} (\bibinfo {year} {1995})},\ \bibinfo {note}
  {publisher: American Physical Society}\BibitemShut {NoStop}%
\bibitem [{\citenamefont {Dudarev}\ \emph {et~al.}(1998)\citenamefont
  {Dudarev}, \citenamefont {Botton}, \citenamefont {Savrasov}, \citenamefont
  {Humphreys},\ and\ \citenamefont
  {Sutton}}]{dudarev_electron-energy-loss_1998}%
  \BibitemOpen
  \bibfield  {author} {\bibinfo {author} {\bibfnamefont {S.~L.}\ \bibnamefont
  {Dudarev}}, \bibinfo {author} {\bibfnamefont {G.~A.}\ \bibnamefont {Botton}},
  \bibinfo {author} {\bibfnamefont {S.~Y.}\ \bibnamefont {Savrasov}}, \bibinfo
  {author} {\bibfnamefont {C.~J.}\ \bibnamefont {Humphreys}},\ and\ \bibinfo
  {author} {\bibfnamefont {A.~P.}\ \bibnamefont {Sutton}},\ }\bibfield  {title}
  {\bibinfo {title} {Electron-energy-loss spectra and the structural stability
  of nickel oxide: {An} {LSDA}+{U} study},\ }\href
  {https://doi.org/10.1103/PhysRevB.57.1505} {\bibfield  {journal} {\bibinfo
  {journal} {Phys. Rev. B}\ }\textbf {\bibinfo {volume} {57}},\ \bibinfo
  {pages} {1505} (\bibinfo {year} {1998})},\ \bibinfo {note} {publisher:
  American Physical Society}\BibitemShut {NoStop}%
\bibitem [{\citenamefont {Artacho}\ \emph {et~al.}(1999)\citenamefont
  {Artacho}, \citenamefont {S?nchez-Portal}, \citenamefont {Ordej?n},
  \citenamefont {Garc?a},\ and\ \citenamefont
  {Soler}}]{artacho_linear-scaling_1999}%
  \BibitemOpen
  \bibfield  {author} {\bibinfo {author} {\bibfnamefont {E.}~\bibnamefont
  {Artacho}}, \bibinfo {author} {\bibfnamefont {D.}~\bibnamefont
  {S?nchez-Portal}}, \bibinfo {author} {\bibfnamefont {P.}~\bibnamefont
  {Ordej?n}}, \bibinfo {author} {\bibfnamefont {A.}~\bibnamefont {Garc?a}},\
  and\ \bibinfo {author} {\bibfnamefont {J.}~\bibnamefont {Soler}},\ }\bibfield
   {title} {\bibinfo {title} {Linear-{Scaling} ab-initio {Calculations} for
  {Large} and {Complex} {Systems}},\ }\href
  {https://doi.org/10.1002/(SICI)1521-3951(199909)215:1<809::AID-PSSB809>3.0.CO;2-0}
  {\bibfield  {journal} {\bibinfo  {journal} {phys. stat. sol. (b)}\ }\textbf
  {\bibinfo {volume} {215}},\ \bibinfo {pages} {809} (\bibinfo {year}
  {1999})}\BibitemShut {NoStop}%
\bibitem [{\citenamefont {Ozaki}(2003)}]{ozaki_variationally_2003}%
  \BibitemOpen
  \bibfield  {author} {\bibinfo {author} {\bibfnamefont {T.}~\bibnamefont
  {Ozaki}},\ }\bibfield  {title} {\bibinfo {title} {Variationally optimized
  atomic orbitals for large-scale electronic structures},\ }\href
  {https://doi.org/10.1103/PhysRevB.67.155108} {\bibfield  {journal} {\bibinfo
  {journal} {Phys. Rev. B}\ }\textbf {\bibinfo {volume} {67}},\ \bibinfo
  {pages} {155108} (\bibinfo {year} {2003})},\ \bibinfo {note} {publisher:
  American Physical Society}\BibitemShut {NoStop}%
\bibitem [{noa()}]{noauthor_openmx_nodate}%
  \BibitemOpen
  \bibfield  {title} {\bibinfo {title} {{OpenMX}},\ }\href@noop {} {\bibinfo
  {journal} {The code OPENMX pseudoatomic basis functions, and pseudopotentials
  are available under terms of the GNU-GPL on a web site,
  http://www.openmx-square.org/}\ }\BibitemShut {NoStop}%
\bibitem [{\citenamefont {Vanderbilt}(1990)}]{vanderbilt_soft_1990}%
  \BibitemOpen
\bibfield  {journal} {  }\bibfield  {author} {\bibinfo {author} {\bibfnamefont
  {D.}~\bibnamefont {Vanderbilt}},\ }\bibfield  {title} {\bibinfo {title} {Soft
  self-consistent pseudopotentials in a generalized eigenvalue formalism},\
  }\href {https://doi.org/10.1103/PhysRevB.41.7892} {\bibfield  {journal}
  {\bibinfo  {journal} {Phys. Rev. B}\ }\textbf {\bibinfo {volume} {41}},\
  \bibinfo {pages} {7892} (\bibinfo {year} {1990})},\ \bibinfo {note}
  {publisher: American Physical Society}\BibitemShut {NoStop}%
\bibitem [{\citenamefont {Morrison}\ \emph {et~al.}(1993)\citenamefont
  {Morrison}, \citenamefont {Bylander},\ and\ \citenamefont
  {Kleinman}}]{morrison_nonlocal_1993}%
  \BibitemOpen
  \bibfield  {author} {\bibinfo {author} {\bibfnamefont {I.}~\bibnamefont
  {Morrison}}, \bibinfo {author} {\bibfnamefont {D.~M.}\ \bibnamefont
  {Bylander}},\ and\ \bibinfo {author} {\bibfnamefont {L.}~\bibnamefont
  {Kleinman}},\ }\bibfield  {title} {\bibinfo {title} {Nonlocal {Hermitian}
  norm-conserving {Vanderbilt} pseudopotential},\ }\href
  {https://doi.org/10.1103/PhysRevB.47.6728} {\bibfield  {journal} {\bibinfo
  {journal} {Phys. Rev. B}\ }\textbf {\bibinfo {volume} {47}},\ \bibinfo
  {pages} {6728} (\bibinfo {year} {1993})},\ \bibinfo {note} {publisher:
  American Physical Society}\BibitemShut {NoStop}%
\bibitem [{\citenamefont {Katsnelson}\ and\ \citenamefont
  {Lichtenstein}(2000)}]{katsnelson_first-principles_2000}%
  \BibitemOpen
  \bibfield  {author} {\bibinfo {author} {\bibfnamefont {M.~I.}\ \bibnamefont
  {Katsnelson}}\ and\ \bibinfo {author} {\bibfnamefont {A.~I.}\ \bibnamefont
  {Lichtenstein}},\ }\bibfield  {title} {\bibinfo {title} {First-principles
  calculations of magnetic interactions in correlated systems},\ }\href
  {https://doi.org/10.1103/PhysRevB.61.8906} {\bibfield  {journal} {\bibinfo
  {journal} {Phys. Rev. B}\ }\textbf {\bibinfo {volume} {61}},\ \bibinfo
  {pages} {8906} (\bibinfo {year} {2000})}\BibitemShut {NoStop}%
\bibitem [{\citenamefont {Antropov}\ \emph {et~al.}(1997)\citenamefont
  {Antropov}, \citenamefont {Katsnelson},\ and\ \citenamefont
  {Liechtenstein}}]{antropov_exchange_1997}%
  \BibitemOpen
  \bibfield  {author} {\bibinfo {author} {\bibfnamefont {V.}~\bibnamefont
  {Antropov}}, \bibinfo {author} {\bibfnamefont {M.}~\bibnamefont
  {Katsnelson}},\ and\ \bibinfo {author} {\bibfnamefont {A.}~\bibnamefont
  {Liechtenstein}},\ }\bibfield  {title} {\bibinfo {title} {Exchange
  interactions in magnets},\ }\href
  {https://doi.org/10.1016/S0921-4526(97)00203-2} {\bibfield  {journal}
  {\bibinfo  {journal} {Physica B: Condensed Matter}\ }\textbf {\bibinfo
  {volume} {237-238}},\ \bibinfo {pages} {336} (\bibinfo {year}
  {1997})}\BibitemShut {NoStop}%
\bibitem [{\citenamefont {Terasawa}\ \emph {et~al.}(2019)\citenamefont
  {Terasawa}, \citenamefont {Matsumoto}, \citenamefont {Ozaki},\ and\
  \citenamefont {Gohda}}]{terasawa_efficient_2019}%
  \BibitemOpen
  \bibfield  {author} {\bibinfo {author} {\bibfnamefont {A.}~\bibnamefont
  {Terasawa}}, \bibinfo {author} {\bibfnamefont {M.}~\bibnamefont {Matsumoto}},
  \bibinfo {author} {\bibfnamefont {T.}~\bibnamefont {Ozaki}},\ and\ \bibinfo
  {author} {\bibfnamefont {Y.}~\bibnamefont {Gohda}},\ }\bibfield  {title}
  {\bibinfo {title} {Efficient {Algorithm} {Based} on {Liechtenstein} {Method}
  for {Computing} {Exchange} {Coupling} {Constants} {Using} {Localized} {Basis}
  {Set}},\ }\href {https://doi.org/10.7566/JPSJ.88.114706} {\bibfield
  {journal} {\bibinfo  {journal} {J. Phys. Soc. Jpn.}\ }\textbf {\bibinfo
  {volume} {88}},\ \bibinfo {pages} {114706} (\bibinfo {year}
  {2019})}\BibitemShut {NoStop}%
\bibitem [{\citenamefont {{P. Blaha}}\ \emph {et~al.}(2002)\citenamefont {{P.
  Blaha}}, \citenamefont {{K. Schwarz}}, \citenamefont {{G. K. H. Madsen}},
  \citenamefont {{D. Kvasnicka}},\ and\ \citenamefont {{J.
  Luitz}}}]{p_blaha_wien2k_2002}%
  \BibitemOpen
  \bibfield  {author} {\bibinfo {author} {\bibnamefont {{P. Blaha}}}, \bibinfo
  {author} {\bibnamefont {{K. Schwarz}}}, \bibinfo {author} {\bibnamefont {{G.
  K. H. Madsen}}}, \bibinfo {author} {\bibnamefont {{D. Kvasnicka}}},\ and\
  \bibinfo {author} {\bibnamefont {{J. Luitz}}},\ }\href@noop {} {\bibinfo
  {title} {{WIEN2K}}} (\bibinfo {year} {2002}),\ \bibinfo {note} {pages: 2201
  Publication Title: J. Phys. Chem. Sol. Volume: 63}\BibitemShut {NoStop}%
\bibitem [{\citenamefont {Smolyanyuk}\ \emph {et~al.}(2024)\citenamefont
  {Smolyanyuk}, \citenamefont {{\v S}mejkal},\ and\ \citenamefont
  {Mazin}}]{smolyanyuk_tool_2024}%
  \BibitemOpen
  \bibfield  {author} {\bibinfo {author} {\bibfnamefont {A.}~\bibnamefont
  {Smolyanyuk}}, \bibinfo {author} {\bibfnamefont {L.}~\bibnamefont {{\v
  S}mejkal}},\ and\ \bibinfo {author} {\bibfnamefont {I.~I.}\ \bibnamefont
  {Mazin}},\ }\href {https://doi.org/10.48550/arXiv.2401.08784} {\bibinfo
  {title} {A tool to check whether a symmetry-compensated collinear magnetic
  material is antiferro- or altermagnetic}} (\bibinfo {year} {2024}),\ \bibinfo
  {note} {arXiv:2401.08784 [cond-mat]}\BibitemShut {NoStop}%
\bibitem [{\citenamefont {Liu}\ \emph {et~al.}(2021)\citenamefont {Liu},
  \citenamefont {Toriyama}, \citenamefont {Cai}, \citenamefont {Zhao},
  \citenamefont {Liu},\ and\ \citenamefont
  {Jeffrey~Snyder}}]{liu_finding_2021}%
  \BibitemOpen
  \bibfield  {author} {\bibinfo {author} {\bibfnamefont {Y.}~\bibnamefont
  {Liu}}, \bibinfo {author} {\bibfnamefont {M.~Y.}\ \bibnamefont {Toriyama}},
  \bibinfo {author} {\bibfnamefont {Z.}~\bibnamefont {Cai}}, \bibinfo {author}
  {\bibfnamefont {M.}~\bibnamefont {Zhao}}, \bibinfo {author} {\bibfnamefont
  {F.}~\bibnamefont {Liu}},\ and\ \bibinfo {author} {\bibfnamefont
  {G.}~\bibnamefont {Jeffrey~Snyder}},\ }\bibfield  {title} {\bibinfo {title}
  {Finding the order in complexity: {The} electronic structure of 14-1-11 zintl
  compounds},\ }\href {https://doi.org/10.1063/5.0068386} {\bibfield  {journal}
  {\bibinfo  {journal} {Applied Physics Letters}\ }\textbf {\bibinfo {volume}
  {119}},\ \bibinfo {pages} {213902} (\bibinfo {year} {2021})}\BibitemShut
  {NoStop}%
\bibitem [{\citenamefont {Siemens}\ \emph {et~al.}(1992)\citenamefont
  {Siemens}, \citenamefont {Del~Castillo}, \citenamefont {Potter},
  \citenamefont {Webb}, \citenamefont {Kuromoto},\ and\ \citenamefont
  {Kauzlarich}}]{siemens_specific_1992}%
  \BibitemOpen
  \bibfield  {author} {\bibinfo {author} {\bibfnamefont {D.~P.}\ \bibnamefont
  {Siemens}}, \bibinfo {author} {\bibfnamefont {J.}~\bibnamefont
  {Del~Castillo}}, \bibinfo {author} {\bibfnamefont {W.}~\bibnamefont
  {Potter}}, \bibinfo {author} {\bibfnamefont {D.~J.}\ \bibnamefont {Webb}},
  \bibinfo {author} {\bibfnamefont {T.~Y.}\ \bibnamefont {Kuromoto}},\ and\
  \bibinfo {author} {\bibfnamefont {S.~M.}\ \bibnamefont {Kauzlarich}},\
  }\bibfield  {title} {\bibinfo {title} {Specific heat of the ternary {Zintl}
  compounds ({Sr14MnBi11} and {Ba14MnBi11})},\ }\href
  {https://doi.org/10.1016/0038-1098(92)90433-A} {\bibfield  {journal}
  {\bibinfo  {journal} {Solid State Communications}\ }\textbf {\bibinfo
  {volume} {84}},\ \bibinfo {pages} {1029} (\bibinfo {year}
  {1992})}\BibitemShut {NoStop}%
\bibitem [{\citenamefont {Gallup}\ \emph {et~al.}(1992)\citenamefont {Gallup},
  \citenamefont {Fong},\ and\ \citenamefont
  {Kauzlarich}}]{gallup_bonding_1992}%
  \BibitemOpen
  \bibfield  {author} {\bibinfo {author} {\bibfnamefont {R.~F.}\ \bibnamefont
  {Gallup}}, \bibinfo {author} {\bibfnamefont {C.~Y.}\ \bibnamefont {Fong}},\
  and\ \bibinfo {author} {\bibfnamefont {S.~M.}\ \bibnamefont {Kauzlarich}},\
  }\bibfield  {title} {\bibinfo {title} {Bonding properties of calcium gallium
  arsenide, {Ca14GaAs11}: a compound containing discrete {GaAs4} tetrahedra and
  a hypervalent {As3} polyatomic unit},\ }\href
  {https://doi.org/10.1021/ic00027a022} {\bibfield  {journal} {\bibinfo
  {journal} {Inorg. Chem.}\ }\textbf {\bibinfo {volume} {31}},\ \bibinfo
  {pages} {115} (\bibinfo {year} {1992})}\BibitemShut {NoStop}%
\end{thebibliography}%

\end{document}